\documentclass[aps,prl,superscriptaddress,showpacs,twocolumn,preprintnumbers,amsmath,amssymb,amsfonts,floatfix]{revtex4-1}

\usepackage{graphicx, color,rotating, textcomp} % Include figure files
\usepackage{dcolumn} % Align table columns on decimal point
\usepackage{bm} % bold math

\begin{document}

\title{Is CrO$_2$ Fully Spin-Polarized? - Analysis of Andreev Spectra and Excess Current}

\author{Tomas~L\"ofwander}
\affiliation{Department of Microtechnology and Nanoscience - MC2,
Chalmers University of Technology, SE-412 96 G\"oteborg, Sweden}

\author{Roland~Grein}
\affiliation{Institut f\"{u}r Theoretische Festk\"{o}perphysik
and DFG-Center for Functional Nanostructures,
Karlsruhe Institute of Technology, D-76128 Karlsruhe, Germany}

\author{Matthias~Eschrig$^{2,}$}
\affiliation{Institut f\"{u}r Theoretische Festk\"{o}perphysik
and DFG-Center for Functional Nanostructures,
Karlsruhe Institute of Technology, D-76128 Karlsruhe, Germany}
\affiliation{Fachbereich Physik, Universit\"at Konstanz, D-78457 Konstanz, Germany}

\date{\today}

\begin{abstract}

  We report an extensive theoretical analysis of point-contact Andreev
  reflection data available in literature on ferromagnetic CrO$_2$. We
  find that the spectra can be well understood within a model of fully
  spin-polarized bands in CrO$_2$ together with spin active scattering
  at the contact. This is in contrast to analyses of the data within
  extended Blonder-Tinkham-Klapwijk models, which lead to a spin
  polarization varying between 50~\% and 100~\% depending on the
  transparency of the interface. We propose to utilize both the
  temperature dependence of the spectra and the excess current at
  voltages above the gap to resolve the spin-polarization in CrO$_2$
  in a new generation of experiments.

\end{abstract}

\pacs{74.55.+v,74.45.+c,72.25.Mk}

\maketitle

Half-metallic ferromagnets are materials with one of the two spin
bands metallic and the other insulating. Their characterization has
attracted great attention, since a fully spin polarized ferromagnetic
material can be very useful for fabricating spin batteries and ideal
magnetic tunnel junctions used in spintronics applications
\cite{pick2001}.  There are only a few materials that are suspected
half-metals \cite{kat08}, one of them being CrO$_2$.

Following early experiments by Soulen {\it et al.} \cite{soulen98} and
by Upadhyay {\it et al.} \cite{Upad}, point-contact Andreev reflection
(PCAR) has been extensively used to probe the spin-polarization of
strong ferromagnets. In this method a nano-sized point contact is
formed by pressing a superconducting tip into the ferromagnetic
material. The conductance-voltage characteristics is recorded and
compared with theory in order to extract the polarization of the
ferromagnet. The key ingredient in these experiments is the
suppression with increased spin polarization of Andreev reflection
processes. Andreev reflection is the scattering event at which an
electron quasiparticle incident from a non-superconducting metal is
retro-reflected as a hole quasiparticle in the opposite-spin
band. Charge conservation is upheld by injection of a Cooper pair into
the superconductor. With increased spin-polarization such
retro-reflection is suppressed.

In recent experiments \cite{keizer06,anwar10}, supercurrent was
observed to flow in long Josephson junctions of CrO$_2$. If CrO$_2$ is
half-metallic, these observations indicate that a conversion of
supercurrent carried by spin singlet Cooper pairs in the
superconductors to supercurrent carried by triplet (equal spin) Cooper
pairs in the halfmetal is taking place. We have \cite{eschrig08}
proposed a conversion mechanism based on spin-active interface
scattering. The question arises, if this model stands the test of
other experiments. Here, we show that spin-active interface scattering
can explain also the PCAR experiments
\cite{soulen98,desisto00,ji01,angu01,angu02,osofsky00,osofsky01,woods04,yates07}
on CrO$_2$ within a model that assumes fully spin-polarized bands.

\begin{figure}[b]
\includegraphics[width=0.8\columnwidth]{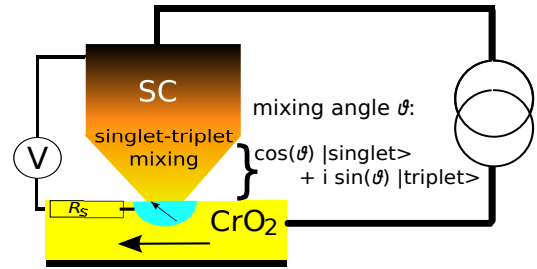}
\caption{Point-contact between a superconductor (SC) and ferromagnetic
CrO$_2$, including a spin-active scattering region. The arrows indicate
a contact effective magnetic moment and the bulk magnetization.}
\label{geometry}
\end{figure}

\begin{table*}[t]
%\begin{ruledtabular}
\begin{tabular}{clcc|ccccc|cccccc}
\hline & & & & extended & BTK & model & & & spin & active & interface & model & &\\
\hline \# & Reference                                 & tip & $T$ [K] &  $P$ [\%]    & $Z$   & $\Delta$ [meV] & $r_s$ & $\chi^2$ & $P$ [\%] & $Z$ & $\vartheta/\pi$ & $\Delta$ [meV] & $r_s$ & $\chi^2$\\
\hline
1.  & Soulen {\it et al.} \cite{soulen98}                & Nb & 1.6   & 66 & 0.90  & 1.0    & 0.05 & 0.34 & 100 & 0.12   & 0.58 & 1.5  & 0      & 0.32\\
2.  & DeSisto {\it et al.} \cite{desisto00}              & Pb & 1.7   & 54  & 1.0    & 1.2    & 0.02 & 0.038 & 100 & 0.28   & 0.44 & 1.35 & 0.05 & 0.065\\
3.  & Ji {\it et al.} \cite{ji01}  Fig.4a                       & Pb & 1.85  & 57 & 1.1   & 1.48   & 0      & 0.067 & 100 & 0.70    & 0.17 & 1.35 & 0.08 & 0.22\\
4.  & Ji {\it et al.} \cite{ji01}  Fig.4b                       & Pb & 1.85 & 72  & 0.87  & 1.43  & 0.05 & 0.025 & 100 & 0.38   & 0.22 & 1.35 & 0.17 & 0.091\\
5.  & Ji {\it et al.} \cite{ji01}  Fig.4c                       & Pb & 1.85  & 94  & 0.50  & 1.0   & 0.32  & 0.090 & 100 & 0.023 & 0.36 & 1.35 & 0     & 0.093\\
6.  & Ji {\it et al.} \cite{ji01}  Fig.4d                        & Pb & 1.85 & 98  & 0       & 0.9    & 0.42  & 0.062 & 100 & 0       & 0.39 & 1.35 & 0.06 & 0.087\\
7.  & Anguelouch \cite{angu01}  Fig.1a & Pb & 1.6    & 63  & 1.3     & 1.35  & 0      & 0.10  & 100 & 0.57    & 0.23 & 1.35 & 0.06 & 0.15\\
8.  & Anguelouch \cite{angu01}  Fig.1b & Pb & 1.6    & 94  & 0.12   & 1.1   & 0.13  & 0.078 & 100 & 0.26   & 0.28 & 1.35 & 0.02 & 0.081\\
9.  & Anguelouch \cite{angu01}  Fig.1c & Pb & 1.6    & 96   & 0.18   & 1.0   & 0.16 & 0.24 & 100 & 0.058  & 0.34 & 1.35 & 0    & 0.22\\
10.& Anguelouch {\it et al.} \cite{angu02}            & Pb & 1.6    & 97  & 0        & 0.94 & 0.29 & 0.27 & 100 & 0.035 & 0.35 & 1.35 & 0    & 0.28\\
11.& Osofsky {\it et al.} \cite{osofsky00}             & Pb & 1.7   & 64   & 1.0    & 1.17  & 0   & 0.19 & 100 & 0.26   & 0.40 & 1.35 & 0     & 0.21\\
12.& Osofsky {\it et al.} \cite{osofsky01}             & Nb & 1.7    & 70  & 1.0     & 1.3   & 0   & 0.21 & 100 & 0.26    & 0.37 & 1.5 & 0    & 0.20\\
\hline
 & Woods {\it et al.} \cite{woods04}              & Sn/Pb & 1.75 & 80       & 0.96   & 0.59/1.2 & 0.28 & -  &  &  &  &  & \\
 & Yates {\it et al.} \cite{yates07}                      & Pb  & 4.2 & 65-100 & 0-1.7  & 0.9-1.3  & -                         & - &  &  &  &  & \\
\hline
\end{tabular}
%\end{ruledtabular}
\caption{Results of non-linear curve-fits of the extended BTK model
(middle set of columns) and the spin-active interface model
(right set of columns) to point-contact Andreev reflection data on CrO$_2$
with superconducting tips of Nb ($\Delta(T=0)=1.5$ meV, $T_c=9.2$ K) and
Pb ($\Delta(T=0)=1.35$ meV, $T_c=7.2$ K). We have obtained improved fits
to the extended BTK model \cite{Mazin} by including a series resistance
$r_s=R_s/R_n$ normalized to the point contact resistance $R_n$ as a fourth
fit parameter \cite{woods04}. The resulting spin polarization $P$,
barrier strength parameter $Z$, and zero temperature gap parameter $\Delta$
are therefor different than found in the original papers. In the spin-active
interface model the bulk quantities are fixed (P=100~\% and $\Delta$ retains
its bulk value), while two interface parameters (barrier strength $Z$
and spin mixing angle $\vartheta$) and the series resistance $r_s$ have been
used as fit parameters. The report in Ref. \cite{yates07} only contains
the fit parameters, but no spectra.}
\label{par}
\end{table*}

The conventional analysis of the PCAR experiments (see
Fig.~\ref{geometry} for the setup) relies upon an extended
Blonder-Tinkham-Klapwijk (BTK) formula \cite{Mazin}, consisting of two
terms. The first term includes Andreev scattering and is the usual BTK
formula \cite{blonder82} for a point contact with an unpolarized
material. The second term is the conductance of a point contact with a
completely spin polarized material for which Andreev reflection is
absent. The two terms are weighted according to the formula $G =
(1-P)G_N + PG_H$, where $P$ is the transport spin polarization. Taking
the limits $V\rightarrow 0$ and $T\rightarrow 0$, the conductance
would reach $G(0)=2(1-P)G_n$ for an ideal contact without
backscattering (unit transparency) since then $G_N\rightarrow 2G_n$
while $G_H=0$ in the whole subgap range. Here, $G_n$ is the
conductance in the non-superconducting state. The BTK model includes
an interface barrier modelled by a delta-function potential quantified
by a dimensionless parameter $Z$, where $Z=0$ corresponds to unit
transparency and $Z\gg 1$ to low transparency. The fit of the
conductance-vs-voltage data results in values for the barrier strength
$Z$ and the polarization $P$. It has often been necessary to also use
the superconducting gap $\Delta$ as fit parameter. Since the gap is a
fit parameter, the gap feature in the spectrum is shifted in an
unpredictable way. A shifted peak position has been attributed to
either a suppressed superconducting order parameter near the contact
(shift to lower voltage) or a spread resistance (shift to higher
voltage). The spread resistance $R_s$ is the resistance of the
material between the contact (the tip) and the voltage probe which
should have been eliminated if a true four-point measurement could
have been set up \cite{woods04}, see Fig.~\ref{geometry}.

It is very unsatisfactory that the existing analyses of the PCAR
experiments
\cite{soulen98,desisto00,ji01,angu01,angu02,osofsky00,osofsky01,woods04,yates07}
with the extended BTK formula have given a wide spectrum of spin
polarizations in CrO$_2$ ranging from 50~\% to 100~\%. In most
experiments the barrier strength is low, with a $Z$ between $0$ and
$2$. This sample to sample variation leads to a most likely spurious
dependence of the polarization (a bulk material property) on $Z$ (an
interface property). It has been argued that the intrinsic
polarization can be obtained by extrapolating the dependence $P(Z)$ to
$Z=0$ (for example Ref.~\cite{angu02}), although this procedure is
questionable since the functional dependence $P(Z)$ is unknown
\cite{woods04}. The suppression of spin polarization from the expected
$P=100~\%$ has in some cases been attributed to the unknown interface
region. Here, we provide a model for this effect.  Another
unsatisfactory feature of the extended BTK model is that the gap
$\Delta$ must be used as fit parameter. Indeed, well characterized
superconducting STM tips display the bulk gap \cite{rod2004}.

In a different experimental set-up \cite{parker02}, Zeeman split
conductance curves of CrO$_2$/Al junctions with fabricated tunnel
barriers were measured with the Meservey-Tedrow technique
\cite{meservey94}. The observed simple linear shift of the spectra
with applied field strongly points to complete spin polarization in
CrO$_2$, in sharp contrast to the picture emerging from the analysis
of the PCAR data with the extended BTK theory.

We give here an alternative interpretation of the PCAR experiments on
CrO$_2$, where we at the outset assume a spin polarization of 100~\%
at the Fermi level and then fit the data with two parameters: besides
$Z$ we utilize the spin-mixing angle $\vartheta$ that describes the
difference in scattering phase factors picked up by electrons of
opposite spin. This spin-mixing leads to a mixture of singlet and
triplet Cooper pairs in the superconductor close to the contact. If
spin-flip scattering is present at the contact, triplet correlations
are induced also in CrO$_2$ \cite{eschrig08}. Spin-flips can be
induced for example by having misaligned moments at the contact, as
depicted by the light blue contact area in Fig.~\ref{geometry} where
the magnetization is misaligned with respect to the bulk CrO$_2$
\cite{misaligned}. When discussing PCAR, this physics is closely
related to a spin-flip Andreev reflection process that leads to
enhanced sub-gap conductance, here simply tuned by the fit-parameter
$\vartheta$. The formula for the conductance relevant for a
half-metallic ferromagnet with a spin-active interface has been
derived within the quasiclassical theory of superconductivity and is
presented as Eqs.~(153)-(157) in Ref.~\cite{eschrig09}.

\begin{figure}[t]
\includegraphics[width=\columnwidth]{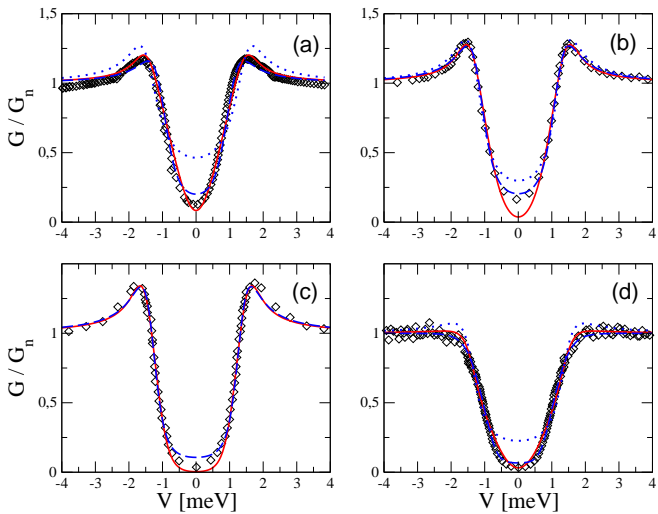}
\caption{Point-contact Andreev reflection data on CrO$_2$ from the
  literature and non-linear curve fits to the theory of
  superconductor-halfmetallic ferromagnet point contact with
  spin-active interface (red solid lines) and the extended BTK model
  (blue dashed lines). The blue dotted lines are fits to the modified
  BTK model without using $\Delta$ as fit parameter. Panels (a)-(d)
  are the data sets 1, 2, 7, and 10 in Table 1.}
\label{fit}
\end{figure}

We have made a fit to all available PCAR data on CrO$_2$ with the
spin-active interface model \cite{modeldetails}. To make the
comparison between the two scenarios consistent, we have also made new
improved fits of the data to the extended BTK model of Mazin {\it et
  al.} \cite{Mazin}, since a variety of extended BTK models have been
used in the literature. The resulting fit parameters are presented in
Table 1 and four representative data fits are shown in Fig.~\ref{fit}.

In Fig.~\ref{fit}(a) we show the data of Soulen {\it et al.}
\cite{soulen98} together with the excellent fit to the spin-active
interface model (red solid line). The blue dotted line is the fit to
the extended BTK model without using the gap as fit
parameter. Clearly, the fit is not good. On the other hand, by letting
$\Delta$ vary freely the fit can be improved except for about 10 data
points in the low-voltage region (blue dashed line, $\Delta$ reduced
by 33~\%). The overall fit, the $\chi^2$ measure, is satisfactory in
this case. In Fig.~\ref{fit}(b) we show the fit to a set of data from
DeSisto {\it et al.} \cite{desisto00}. In this case the modified BTK
fit can be made excellent by reducing $\Delta$ by 10~\%. Both models
give satisfactory $\chi^2$ measures, although three data points at low
voltage are not perfectly fitted by the spin-active model.  In
Figs.~\ref{fit}(c)-(d) we show two other representative fits to both
models. Again, in (d) the gap had to be reduced by about 30~\% to
improve the fit to the modified BTK model.

\begin{figure}[t]
\includegraphics[width=\columnwidth]{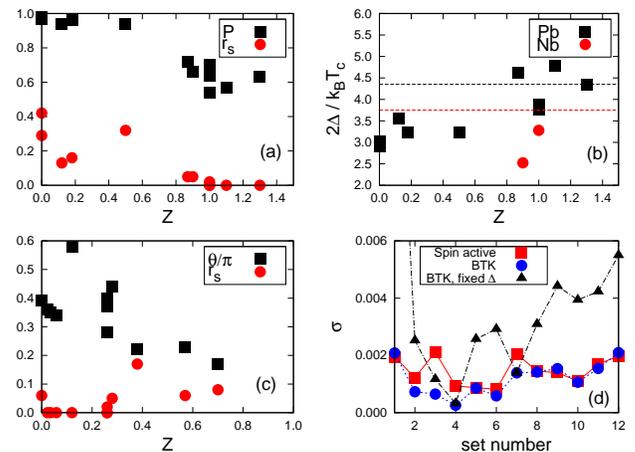}
\caption{Correlations between fit parameters. (a) Spin polarization
  $P$ and series resistance $r_s$ versus the barrier strength $Z$ in
  the modified BTK model. (b) The superconducting gap $\Delta$ versus
  $Z$ in the modified BTK model. (c) Spin mixing angle $\vartheta$ and
  $r_s$ versus $Z$ in the spin-active interface model. (d) Comparison
  of the fitting function $\chi^2$ divided by the number of data
  points $N$ for the two models.}
\label{correlations}
\end{figure}

From the obtained fit parameters presented in Table~\ref{par}, we can
draw several conclusions. The main finding of this paper is that from
a non-linear curve fit ($\chi^2$) perspective, all data sets can be
fit well with the spin-active interface model with 100~\% spin
polarization at the Fermi level of CrO$_2$. This provides a
complementary picture to the one emerging from the traditional
extended BTK model fit. This is also reflected in the same order of
magnitudes in the calculated average variances $\sigma=\chi^2/N$
(where N is the number of data points in each set) for the two models,
shown in Fig.~\ref{correlations}(d). The spin-mixing angle, here
quantifying spin-flip Andreev reflection at the interface, is varying
between samples but is suppressed in high-$Z$ junctions. This is
consistent with the picture of the origin of the spin-mixing effect at
interfaces to strong ferromagnets \cite{eschrig08}, where the minority
spin electrons gain an additional phase compared with majority spins
during reflection inside the classically forbidden region in the
half-metallic ferromagnet. As the barrier between the materials is
enhanced, this effect is reduced.

The wide spread of the spin polarization $P$ resulting from the
extended BTK model (50~\% to 100~\%) is unsatisfactory, since $P$ is a
bulk property. We find a correlation between spin-polarization $P$ and
the barrier strength $Z$, see Fig.\ref{correlations}(a), where (as
reported before) $P$ appears to approach 100~\% as $Z\rightarrow
0$. We have utilized the (normalized to the contact resistance) spread
resistance $r_s=R_s/R_n$ as fit parameter in both models, although it
plays a more important role for the fit with the extended BTK
model. The variation of $r_s$ with barrier strength is shown in
Fig.~\ref{correlations}(a) and (c). Only in one reference has the
spread resistance been discussed seriously (Woods {\it et al.}
\cite{woods04}). However, because we found an uncertainty concerning
their data we do not present a fit here \cite{inconsistency}. It is
crucial to use $\Delta$ as fit parameter in the extended BTK model,
while such variation does not improve the fits to the spin-active
interface model. It is unclear why the (bulk) $\Delta$ should vary so
much [see Fig.~\ref{correlations}(b)], in contrast to other
experiments with superconducting STM tips \cite{rod2004}. All point
contacts appear to be highly transparent with a small barrier strength
parameter $Z<1$. This is consistent with a Fermi velocity mismatch,
which for Pb/CrO$_2$ was estimated in Ref.~\cite{yates07} to give
$Z\approx 0.26$. But this phenomenological parameter should include a
range of effects causing mismatch between the materials.

We are able to fit the experimental data with three parameters
describing the interface ($\vartheta$ and $Z$) or the geometry
($r_s$), while the extended BTK model relies for the same spectra on
four fit parameters among which two ($P$ and $\Delta$) pertains to
bulk properties, one to the interface ($Z$), and one to the geometry
($r_s$). The part of the spectra hardest to fit to either model is the
low-voltage region. In this region there are typically much less data
points than in the high voltage region [for example Fig.\ref{fit}(b)
and (c)]. This typically happens in a current bias set-up, which is
not ideal for PCAR. Our fits could be further improved if we would
allow for broadening in the form of a convolution with a Gaussian (as
used in Ref.~\cite{yates07}), which would describe e.g. voltage
fluctuations. Since we do not know all experimental uncertainties we
leave this question open for future experiments.

\begin{figure}[t]
\includegraphics[width=\columnwidth]{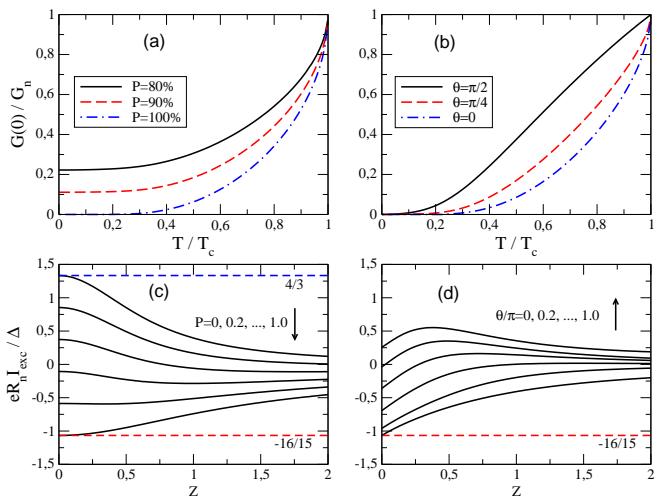}
\caption{Temperature dependence of the zero-voltage conductance as
  predicted by (a) the modified BTK model and (b) the spin-active
  interface model. Excess current at $V\gg\Delta/e$ versus barrier
  strength parameter $Z$ as predicted by (c) the modified BTK model
  and (d) the spin-active interface model.}
\label{G0_vs_T}
\end{figure}

We would like to point out a few details of importance for future PCAR
experiments, which may shed more light on the properties of
CrO$_2$. Thermal smearing is important, since it gives a considerable
increase of $G(V=0)$ as compared with the $T\rightarrow 0$ limit at
the temperatures used in the experiments.  For the spin-active
interface model, $G(0)\rightarrow 0$ as $T\rightarrow 0$ independently
of the barrier strength. This is a unique feature that has not been
fully explored experimentally. In our model this is a result of
vanishing spectral current $j_{\varepsilon}=0$ at the Fermi energy
$\varepsilon=0$. In contrast, in the extended BTK model, $G(0)$
saturates at a value given by the polarization and barrier strength,
see Fig.~\ref{G0_vs_T}(a)-(b). Thus, the temperature dependence of
$G(0)$ in a well-defined voltage bias set-up can be used as a
consistency check between experiment and theory.

Another quantity that has not been explored experimentally so far is
the excess current, formally defined as
$I_{exc}=\lim_{V\rightarrow\infty}\left[I(V)-I_n(V)\right]$, where
$I_n(V)$ is the current in the normal state ($=R_nV$ according to
Ohm's law). In certain limits, the excess current can be computed
analytically although the formulas are rather cumbersome. We present
the excess currents predicted by the two models in
Fig.~\ref{G0_vs_T}(c)-(d). A measurement of the excess current in
addition to the PCAR spectrum can be used to pin down one of the fit
parameters (or as consistency check) in future experiments.

In conclusion, a number of PCAR spectra of CrO$_2$ have been presented
in the literature where, by comparing the data to extended BTK models,
a putative spin polarization between 50~\% and 100~\% has been
extracted. This is in contrast to Zeeman split conductance
measurements where 100~\% polarization was found. We have provided an
alternative view of the PCAR data, where the spin polarization is
100~\%, but the scattering at the contact is spin-active.

%\vspace{-0.5cm}


\begin{thebibliography}{27}
\bibitem{misaligned} Here we report results for a misalignment angle
  $\alpha=\pi/2$. The resulting fits are qualitatively the same for
  $\alpha>0.3\pi$.
\bibitem{modeldetails} Our model includes a variation of the
  spin-mixing angle with impact angle $\theta_p$ in accordance with a
  $\delta$-function potential of strength $Z$:
  $\vartheta(\theta_p)=\vartheta(0)\cos^2(\theta_p)$.
\bibitem{inconsistency} The data shown in Ref.~\cite{woods04}, Fig.~4,
  is presented as being for a Sn/CrO$_2$ contact. However, using
  $T=1.75$ K and $T_c^{Sn}=3.7$ K or $T_c^{Pb}=7.2$ K in the thermally
  smeared version of the formulas presented in Ref.~\cite{Mazin} we
  could reproduce the curve in Fig.~4(a) of Ref.~\cite{woods04} only
  by assuming a Pb counter electrode.
\end{thebibliography}
\end{document}